\documentclass[prb,twocolumn]{revtex4}%

\usepackage{graphicx}
\usepackage{amsmath}
\usepackage{ulem}
\usepackage{color}

\newcommand{\be}{\begin{equation}}
\newcommand{\ee}{\end{equation}}
\newcommand{\bea}{\begin{eqnarray*}}
\newcommand{\eea}{\end{eqnarray*}}
\newcommand{\bean}{\begin{eqnarray}}
\newcommand{\eean}{\end{eqnarray}}

\begin{document}

\draft
\title
{\bf Thermoelectric properties of finite two-dimensional quantum
dot arrays with band-like electronic states}

\author{David M T Kuo }
\address{Department of Electrical Engineering and Department of Physics, National Central
University, Chungli, 320 Taiwan}



\date{\today}

\begin{abstract}
The thermal power ($PF=S^2G_e$) depends on the Seebeck coefficient
($S$) and electron conductance ($G_e$). The enhancement of $G_e$
will unavoidably suppress $S$ because they are closely related. As
a consequence, the optimization of $PF$ is extremely difficult.
Here, we theoretically investigated the thermoelectric properties
of two-dimensional quantum dot (QD) arrays with carriers injected
from electrodes. The Lorenz number of 2D QD arrays in the resonant
tunneling procedure satisfies the Wiedemann-Franz law, which
confirms the formation of minibands. When the miniband center is
far away from the Fermi level of the electrodes, the electron
transport is in the thermionic-assisted tunneling procedure
(TATP). In this regime, $G_e$ in band-like situation and $S$ in
atom-like situation can happen simultaneously. We have
demonstrated that the enhancement of $G_e$ with an increasing
number of electronic states will not suppress $S$ in the TATP.
\end{abstract}

\maketitle

\section{Introduction}
The semiconductor quantum dots (QDs) resulting from the quantum
confinement of heterostructures exhibit atom-like discrete
electron energy levels. High-efficiency single-QD devices show the
functionalities of low electrical and optical power outputs. These
single-QD devices include single electron
transistors[\onlinecite{Guo}-\onlinecite{Kuba}], single photon
sources[\onlinecite{Michler}-\onlinecite{Chang}], single photon
detectors[\onlinecite{Gustavsson}] and single electron heat
engines[\onlinecite{Josefsson}]. Some applications of QD devices
require both high efficiency and significant output power.
Therefore, one needs QD solids that can retain the size tunable
properties of the QDs while exhibiting the band transport
characteristic of bulk semiconductors.[\onlinecite{Kagan}]
Although much effort has been devoted to producing such QD solids,
studies of the thermoelectric properties of such 2D QD arrays have
been lacking.[\onlinecite{Harman},\onlinecite{Talgorn}]

Designing a thermoelectric material with a high figure of merit
($ZT$) and optimized power output is under
pursuit.[\onlinecite{Kuo1}-\onlinecite{Kuo2}] The dimensionless
figure of merit $ZT=S^2G_eT/\kappa$ depend on the Seeback
coefficient ($S$), electrical conductance ($G_e$) and thermal
conductance ($\kappa$) of the material. Although 1D QD arrays have
very high $ZT$ values, there exist many limitations in the
implementations of thermoelectric devices.[\onlinecite{Kagan}] 2D
and 3D QD arrays are required for realistic applications. The
$\kappa$ of a 2D QD array is smaller than that of bulk
material.[\onlinecite{ChenG}] This low dimensional system has the
potential to realize high $ZT$ values.[\onlinecite{Chen}]
Therefore, it is desirable to investigate the power factor
($PF=S^2G_e$) of 2D QD arrays, which directly affects the
electrical power output. The enhancement of $G_e$ calls for a
large number of electronic states (band-like). However a large $S$
value occurs in dilute electronic states (atom-like). Therefore,
enhancing one of these physical quantities will unavoidably
suppress the other. This study theoretically investigated the
thermoelectric properties of a finite 2D QD array coupled to
electrodes, as shown in Fig. 1. The electrons of the QD array are
injected from the electrodes.[\onlinecite{Mahan}] We demonstrated
that $G_e$ in band-like situation and $S$ in atom-like situation
can happen simultaneously when the miniband center of a 2D QD
array remains a certain distance from the Fermi level of the
electrodes. These results will improve the thermoelectric
performance of 2D materials such as $SnSe$ and
$MoS_2$.[\onlinecite{Zhao}-\onlinecite{Fan}]

\begin{figure}[h]
\centering
\includegraphics[trim=2.5cm 0cm 2.5cm 0cm,clip,angle=-90,scale=0.3]{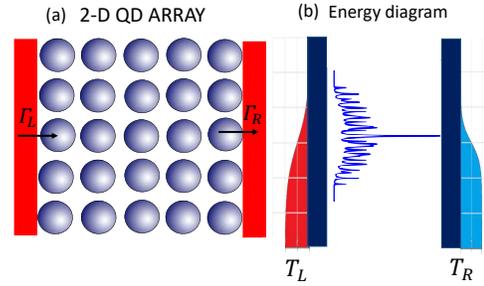}
\caption{(a) Schematic diagram of a two dimensional (2D) quantum
dot (QD) array coupled to electrodes. $\Gamma_{L}$ ($\Gamma_R$)
denotes the tunneling rate of the electrons between the left
(right) electrode and the leftmost (rightmost) QDs. Energy diagram
of a 2D QD array coupled to electrodes with different equilibrium
temperatures ($T_L$ and $T_R$). }
\end{figure}

\section{Formalism}

To model the thermoelectric properties of a 2D QD array connected
to the electrodes, the Hamiltonian of the system shown in Fig. 1
is given by $H=H_0+H_{QD}$,[\onlinecite{Haug}] where
\begin{small}
\begin{eqnarray}
H_0& = &\sum_{k,\sigma} \epsilon_k
a^{\dagger}_{k,\sigma}a_{k,\sigma}+ \sum_{k,\sigma} \epsilon_k
b^{\dagger}_{k,\sigma}b_{k,\sigma}\\ \nonumber
&+&\sum_{\ell}^{N_y}\sum_{k,\sigma}
V^L_{k,\ell,j}d^{\dagger}_{\ell,j,\sigma}a_{k,\sigma}
+\sum_{\ell}^{N_y}\sum_{k,\sigma}V^R_{k,\ell,j}d^{\dagger}_{\ell,j,\sigma}b_{k,\sigma}+H.c.
\end{eqnarray}
\end{small}
The first two terms of Eq.~(1) describe the free electron gas in
the left and right electrodes. $a^{\dagger}_{k,\sigma}$
($b^{\dagger}_{k,\sigma}$) creates  an electron of momentum $k$
and spin $\sigma$ with energy $\epsilon_k$ in the left (right)
electrode. $V^L_{k,\ell,j}$ ($V^R_{k,\ell,j}$) describes the
coupling between the left (right) lead with its adjacent QD in the
$\ell$th row, which counts from 1 to $N_y$.
\begin{small}
\begin{eqnarray}
H_{QD}&= &\sum_{\ell,j,\sigma} E_{\ell,j}
d^{\dagger}_{\ell,j,\sigma}d_{\ell,j,\sigma}\\ \nonumber&+&
\sum_{\sigma}\sum_{\ell 1,\ell 2}^{N_y}\sum_{j1,j2}^{N_x} t_{\ell
1,\ell 2, j1, j2} d^{\dagger}_{\ell 1,j1,\sigma} d_{\ell
2,j2,\sigma}+H.c,
\end{eqnarray}
\end{small}
\begin{equation}
t_{\ell 1,\ell 2,j1, j2}= \{ \begin{array}{ll} -t_{y} &
if~j1=j2,  |\ell 1-\ell 2|=1\\
-t_{x} & if~\ell 1=\ell 2, |j1-j2|=1
\end{array},
\end{equation}
where { $E_{\ell,j}$} is the energy level of QD  in the
${\ell}$-th row and $j$-th column. The spin-independent $t_{\ell
1, \ell 2, j1, j2}$ describes the electron hopping strength, which
is limited to the nearest neighboring sites. $d^{\dagger}_{\ell
1,j1,\sigma} (d_{\ell 2,j2,\sigma})$ creates (destroys) one
electron in the QD at the $\ell$th row and $j$th column. If the
wave functions of the electrons in each QD are localized, the
electron Coulomb interactions are strong. Their effects on
electron transport are significant in the scenario of weak hopping
strengths.[\onlinecite{Kuo3}] On the other hand, the wave
functions of the electrons are delocalized in the scenario of
strong hopping strengths to form minibands; hence their weak
electron Coulomb interactions can be ignored.

To study the transport properties of a 2D QD array junction
connected to electrodes, it is convenient to use the
Keldysh-Green's function
technique[\onlinecite{Haug},\onlinecite{Meir}]. Electron and heat
currents leaving electrodes can be expressed as
\begin{eqnarray}
J &=&\frac{2e}{h}\int {d\varepsilon}~
T_{LR}(\varepsilon)[f_L(\varepsilon)-f_R(\varepsilon)],
\end{eqnarray}
and
\begin{eqnarray}
& &Q_{e,L(R)}\\ &=&\frac{\pm 2}{h}\int {d\varepsilon}~
T_{LR}(\varepsilon)(\varepsilon-\mu_{L(R)})[f_L(\varepsilon)-f_R(\varepsilon)]\nonumber
\end{eqnarray}
where
$f_{\alpha}(\varepsilon)=1/\{\exp[(\varepsilon-\mu_{\alpha})/k_BT_{\alpha}]+1\}$
denotes the Fermi distribution function for the $\alpha$-th
electrode, where $\mu_\alpha$  and $T_{\alpha}$ are the chemical
potential and the temperature of the $\alpha$ electrode. $e$, $h$,
and $k_B$ denote the electron charge, the Planck's constant, and
the Boltzmann constant, in that order. $T_{LR}(\varepsilon)$
denotes the transmission coefficient of a 2D QD array connected to
electrodes, which can be solved by the formula $
T_{LR}(\varepsilon)=4Tr[\hat{\Gamma}_{L}\hat{G}^{r}_{D,A}(\varepsilon)\hat{\Gamma}_{R}\hat{G}^{a}_{D,A}(\varepsilon)]$,
where the matrix of tunneling rates ($\hat{\Gamma}_L$ and
$\hat{\Gamma}_R$) and Green's functions
($\hat{G}^{r}_{D,A}(\varepsilon)$ and
$\hat{G}^{a}_{D,A}(\varepsilon)$) can be constructed by
coding.[\onlinecite{Kuo4}]

The electrical conductance ($G_e$), Seebeck coefficient ($S$) and
electron thermal conductance ($\kappa_e$) can be evaluated by
using Eqs. (4) and (5) with a small applied bias $\Delta
V=(\mu_L-\mu_R)/e$ and cross-junction temperature difference
$\Delta T=T_L-T_R$. We obtain these thermoelectric coefficients
$G_e=e^2{\cal L}_{0}$, $S=-{\cal L}_{1}/(eT{\cal L}_{0})$ and
$\kappa_e=\frac{1}{T}({\cal L}_2-{\cal L}^2_1/{\cal L}_0)$. ${\cal
L}_n$ is given by
\begin{equation}
{\cal L}_n=\frac{2}{h}\int d\varepsilon~
T_{LR}(\varepsilon)(\varepsilon-E_F)^n\frac{\partial
f(\varepsilon)}{\partial E_F},
\end{equation}
where $f(\varepsilon)=1/(exp^{(\varepsilon-E_F)/k_BT}+1)$ is the
Fermi distribution function of electrodes at equilibrium
temperature $T$.

\section{ Results and discussion}
According to Eq. (6), transmission coefficient plays a significant
role for electron transport between the electrodes. To illustrate
the electronic states of finite 2D QD array, we have calculated
and shown in Fig. 2 transmission coefficient $T_{LR}(\varepsilon)$
as a function of $\varepsilon$ for different tunneling rates
($\Gamma_{L(R),\ell,j}(\varepsilon)=2\pi\sum_{k}
|V^{L(R)}_{k,\ell,j}|^2
\delta(\varepsilon-\varepsilon_k)=\Gamma_t$). A square lattice
with homogenous electron hopping strengths $t_x=t_y=t_c=6\Gamma_0$
and site-independent QD energy level $E_{\ell,j}=E_0=E_F$ has been
considered in the calculation of $T_{LR}(\varepsilon)$. All
physical parameters are in units of $\Gamma_0$. In Fig. 2(a),
$T_{LR}(\varepsilon)$ reveals the tunneling probability of the
electrons of the electrodes through the electronic states of 2D QD
array, those energy is described by
$\varepsilon=E_0-2t_c(cos(\frac{n_x\pi}{N_x+1})+
cos(\frac{n_y\pi}{N_y+1}))$, where $n_x=1,2,..N_x$ and
$n_y=1,2,..N_y$.  Because the QD array is connected to the
electrodes, these electronic states have inhomogeneous broadening.
They are also restricted within the range between $-4t_c$ and
$4t_c$. We can tune the distribution of electronic states by
changing $N$, $t_c$ and $\Gamma_t$.

\begin{figure}[h]
\centering
\includegraphics[angle=-90,scale=0.3]{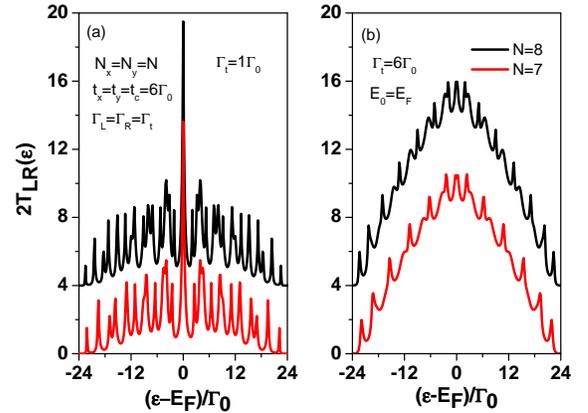}
\caption{Transmission coefficient $T_{LR}(\varepsilon)$ as a
function of $\varepsilon$ for different $N$ values  at
$t_x=t_y=t_c=6\Gamma_0$ and $E_0=E_F$. Diagrams (a) and  (b)
consider tunneling rates $\Gamma_L=\Gamma_R=\Gamma_t=1\Gamma_0$
and $\Gamma_t=6\Gamma_0$, respectively. To prevent the curves from
overlapping each other, we shifted the curve of N = 8.}
\end{figure}

From Eqs. (4) and (5), the maximum electron current and heat
current occur at $T_{LR}(\varepsilon)$ with the maximum area. The
authors of Ref. [\onlinecite{Kuo4}] proved two results: the
maximum area of $T_{LR}(\varepsilon)$ can be reached at the
condition of $\Gamma_t=t_c$ and the maximum area increases with
increasing $N$, as seen in Fig. 2(b). Note that the 2D
tight-binding electronic states show the Van Hove singularity in
the density of states (DOS) as $N \rightarrow \infty$ (DOS
diverges at $E_0$). At zero temperature, the electrical
conductance is given by the transmission coefficient
$G_e=\frac{2e^2}{h}T_{LR}(E_F)$. We now clarify how the electronic
states influence the thermoelectric coefficients of a finite 2D QD
array.

Fig. 3 shows the calculated $G_e$, $\kappa_e$ and Lorenz number
($L_0=\kappa_e/(TG_e)$ at functions of the QD energy level for
various values of $\Gamma_t$ at $k_BT=1\Gamma_0$, $t_c=6\Gamma_0$
and $N=8$. As a result of temperature effect ($\frac{\partial
f(\varepsilon)}{\partial E_F}$), the electronic states shown in
Fig. 2(a) can not be resolved in Fig. 3(a). It is not easy to
justify a finite 2D QD array in the band-like or molecule-like
situation from $G_e$ at finite temperature, especially at high
temperatures. The curves of $\kappa_e$ in Fig. 3(b) are similar to
those of $G_e$. According to the Wiedemann-Franz law,
$L_0/(k^2_B/e^2)=\frac{\pi^2}{3}$ is a temperature-independent
quantity. In Fig. 3(c), the $L_0$ curve corresponding to
$\Gamma_t=6\Gamma_0$ is approximately $\pi^2/3$. For comparison,
we also add the calculated $G_e$, $\kappa_e$ and $L_0$ for 1D QD
array with $N_x=100$ and $t_c=\Gamma_t=12\Gamma_0$ (the band width
of $48\Gamma_0$ in this 1D miniband). As seen in Fig. 3(c) , 1D QD
array yields a Lorenz number
$L_0=\frac{k^2_B}{e^2}\frac{\pi^2}{3}$ between $-10\Gamma_0 \le
\Delta \le 10\Gamma_0$. This can be regarded as a manifested
band-like transport.

\begin{figure}[h]
\centering
\includegraphics[angle=-90,scale=0.3]{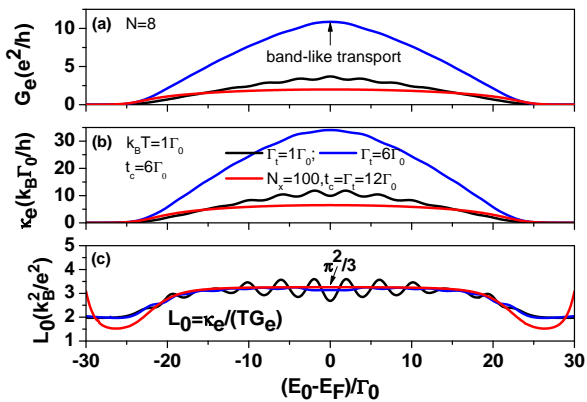}
\caption{(a) Electrical conductance $G_e$, (b) electron heat
conductance $\kappa_e$ and (c) Lorenz number
($L_0=\kappa_e/(TG_e)$) as functions of $\Delta=E_0-E_F$ for
various $\Gamma_t$ values  at $k_BT=1\Gamma_0$, $t_c=6\Gamma_0$
and $N=8$. The red curves correspond to a 1D QD array with
$N_x=100$, $t_c=\Gamma_t=12\Gamma_0$ and $k_BT=1\Gamma_0$.}
\end{figure}

Furthermore, we have calculated $G_e$, $\kappa_e$ and $L_0$ for 2D
QD array with $N=8$ as functions of temperature for different
$\Gamma_t$ values at $E_0=E_F$ and $t_c=6\Gamma_0$ in Fig. 4. The
red curves are the results of the 1D QD array corresponding to
those of Fig. 3. One-dimensional QD arrays have a
temperature-independent $G_e$ and a linear temperature-dependent
$\kappa_e$. As a consequence, $L_0=\kappa_e/(TG_e)$ leads to a
temperature-independent behavior. According to the results of
Figs. 3 and 4, the thermoelectric properties of 2D QD arrays with
$t_c=6\Gamma_0$, $\Gamma_t=6\Gamma_0$ and $N=8$ are very similar
to those of 1D QD arrays with minibands. We deduce that finite 2D
QD arrays have band-like characteristics when
$t_c=\Gamma_t=6\Gamma_0$ and $N=8$.

\begin{figure}[h]
\centering
\includegraphics[angle=-90,scale=0.3]{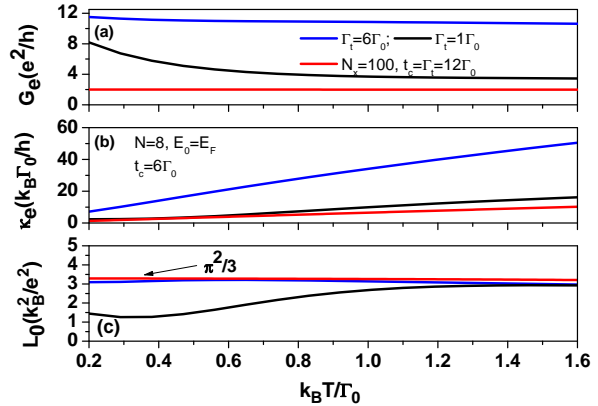}
\caption{(a) Electrical conductance, (b) electron heat conductance
and (c) Lorenz number  as functions of temperature for various
$\Gamma_t$ values at $E_0=E_F$, $t_c=6\Gamma_0$ and $N=8$. The red
curves correspond to those of Fig. 3 with $E_0=E_F$.}
\end{figure}

Because many thermoelectric devices operate at high temperatures,
it is important to examine the power factor of 2D QD arrays in
this regime. In Figs. 2-4 we have focused on the electron
transport in resonant tunneling procedure (RTP) in which the
Seebeck coefficient is very small. To obtain a large $PF$ value at
high temperature, we considered the electron transport in the
thermionic-assisted tunneling procedure (TATP) where the band
center ($E_0$) is far away from the Fermi level $E_F$ in the
electrodes. In Fig. 5, we have calculated $G_e$, $S$, $PF$ and
$L_0$ as functions of temperature for various $t_c$ values at
$E_0-E_F=30\Gamma_0$. Because the maximum $T_{LR}(\varepsilon)$
area occurs at $\Gamma_t=t_c$, we adopted this condition for all
the subsequent steps. As seen in Fig. 5(a), $G_e$ is vanishingly
small at low temperature due to the electronic states of 2D QD
arrays being kept a certain distance from the $E_F$. The
enhancement of $G_e$ with increasing temperature is a typical
characteristic arising from the TATP. To understand the
temperature behavior of $G_e$ at $\Gamma_t=t_c=1\Gamma_0$, we have
the expression of $G_{e,atom}=\frac{e^2}{h} \frac{\pi \Gamma_t}
{2k_BT cosh^2((E_0-E_F)/(2k_BT))}$ when the transmission
coefficient is approximated as
$T_{LR}(\varepsilon)=4\Gamma^2_t/((\varepsilon-E_0)^2+(2\Gamma_t)^2)$
in Eq. (6). In addition, $S_{atom}=-\Delta/T=-(E_0-E_F)/T$, which
explains the behavior of $S$ at $t_c=1\Gamma_0$ and $k_BT \ge
2\Gamma_0$ in Fig. 5(b). Although $G_e$ is highly enhanced with
increasing $t_c$, $S$ is not so sensitive to $t_c$ for $k_BT \ge
10 \Gamma_0$. This explains why the trend of maximum $PF$ for
$t_c$ shown in Fig. 5(c) is determined by $G_e$. In Fig. 5(d),
three $L_0$ curves violate the Wiedemann-Franz law. Note that
$\Gamma_t=t_c=6\Gamma_0$ provides the band-like characteristic
(see Fig. 3(c)).

\begin{figure}[h]
\centering
\includegraphics[angle=-90,scale=0.3]{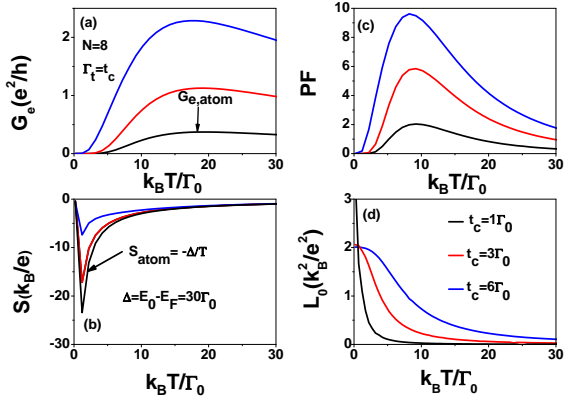}
\caption{(a) Electrical conductance, (b) Seeback coefficient, (c)
power factor ($PF=S^2G_e$) and (d) Lorentz number  as functions of
$T$ for various $t_c$ values at $E_0-E_F=30\Gamma_0$ and $N=8$.
Meanwhile, we have adopted $\Gamma_t=t_c$.}
\end{figure}

Fig. 6 shows the calculated $G_e$, $S$ and $PF$ as functions of
$E_0$ for various $t_c$ values at $k_BT=25\Gamma_0$ to reveal the
effect of the band center. As seen in Fig. 6(a), $G_e$ has a
maximum value when the band center ($E_0$) is located at $E_F$.
The Seeback coefficients in Fig. 6(b) are zero at $E_0=E_F$. It is
attributed to the symmetrical distribution of the electrons and
holes on the electronic states of the 2D QD array. Here, the holes
are defined as the empty states below $E_F$. For
$k_BT=25\Gamma_0$, the Seebeck coefficients are well described by
$S_{atom}=-\Delta/T$. We find that the maximum values of $PF$
occur near $\Delta=60\Gamma_0$, as indicated in Fig. 6(c). When
approaching the atomic limit ($t_c \rightarrow 0$), one can prove
that the optimization of $PF$ is given by $\Delta/k_BT=2.4$. The
results of Fig. 6(c) imply that 2D QD arrays with minibands
($t_c=6\Gamma_0$) preserve the atomic thermoelectric properties
when the band center is far away from the $E_F$ of the electrodes.
In Fig. 2(b), $T_{LR}(\varepsilon)$ depends on $N$. Therefore, we
add in Fig. 6 the curves with triangle marks for $N=7$ and
$t_c=6\Gamma_0$. From the curves of $N=7$ and $N=8$, we see that
the enhancement of $G_e$ resulting from the increase of electronic
states does not suppress $S$. It is worthy noting that a single 1D
QD array does not exist such a behavior. We reinvestigate $PF$ as
functions of $t_c$ for $N=7,8$ at $\Delta=60\Gamma_0$ and
$k_BT=25\Gamma_0$ in Fig. 6(d). $PF$ is a linear function of $t_c$
as $t_c\le 6\Gamma_0$. Meanwhile, the maximum $PF$ is given by
$t_c=\Delta/4$. The red curves represent the case where $t_y=0$ to
clarify the geometer effects. When $t_c \le 6\Gamma_0$, the
geometry effects can be ignored.

Because $T_{LR}(\varepsilon)$ lacks an analytical form, it is not
easy to illustrate the complex behavior of $PF$ shown in Fig.
6(d). If we make the assumption that minibands have homogenous
electronic states and consider the square-form
$T_{LR}(\varepsilon)$ given by
\begin{equation}
T_{LR}(\varepsilon)=\left\{ \begin{array}{ll}
N_y& \mbox {if $-2t_c\le \varepsilon-E_0 \le 2t_c,$}\\
0 &\mbox{otherwise}\end{array} \right.
\end{equation}
the analytical forms of $G_e$ and $S$ can be derived as
\begin{equation}
G_e=\frac{2e^2N_y}{h}~(tanh(y_1)-tanh(y_2))
\end{equation}
and
\begin{equation}
S=\frac{2ek_BN_y}{h}~\frac{(S_1(y_1)-S_2(y_2))}{G_e}
\end{equation}
where $S_i(y_i)=y_i~tanh(y_i)-log(cosh(y_i))$,
$y_1=\frac{\Delta+2t_c}{2k_BT}$ and
$y_2=\frac{\Delta-2t_c}{2k_BT}$.  In Fig 6(d), the blue curves are
calculated by using Eqs. (8) and (9). Because Eq. (7) considers
$N_y$ 1D QD arrays with homogenous electric states, it is expected
that the $PF$ given by Eqs. (8) and (9) is overestimated. However,
Eqs. (8) and (9) provide a clear picture that the enhancement of
$PF$ follows the enhancement of $G_e$. Meanwhile, $S\approx
S_{atom}$ as long as $\frac{t_c}{k_BT} < 0.25$ and
$\frac{\Delta}{k_BT} \ge 2.4$. We deduce that $G_e$ in a band-like
transport situation and $S$ in an atomic-like situation can
coexist for finite 2D QD arrays with $t_c=6\Gamma_0$ and
$\Delta=60\Gamma_0$ at $k_BT=25\Gamma_0$. If we set
$\Gamma_0=1~meV$, then our analysis in Fig. 6 becomes a very
useful guideline for thermoelectric devices operated at room
temperature.

\begin{figure}[h]
\centering
\includegraphics[angle=-90,scale=0.3]{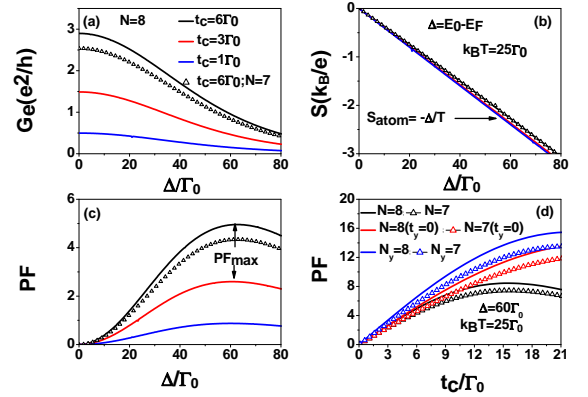}
\caption{(a) Electrical conductance, (b) Seebeck coefficient and
(c) power factor as functions of $E_0$ for different $t_c$ values
at $N=8 $ and $k_BT=25\Gamma_0$. The curves with black triangle
marks correspond to the case of $N=7$ and $t_c=6\Gamma_0$.(d) $PF$
as functions of $t_c$ for $\Delta=60\Gamma_0$ and
$k_BT=25\Gamma_0$. The red curves correspond to $t_y=0$. The blue
curves are calculated using Eq. (7).}
\end{figure}

\section{Conclusion}
We have theoretically investigated the thermoelectric properties
of 2D QD arrays. In RTP, the Lorenz number with a value near
$\pi^2/3$ and a temperature-independent behavior demonstrates that
2D QD arrays with $t_c=\Gamma_t=6\Gamma_0$ and $N=8$ indeed form
minibands. When this miniband center is far away from the Fermi
level of the electrodes, TATP dominates the electron transport
between the electrodes and $L_0$ violates the Wiedemann-Franz law.
In TATP, $G_e$ is enhanced as the number of electronic states
increases, whereas the $S$ values remain in an atom-like
situation. This is a remarkable property that would lead to a
high-efficiency thermoelectric devices made of QDs with large
electrical power output. This interesting phenomenon exists not
only for 2D QD arrays with square-lattices but also
triangular-lattices, which will be reported in elsewhere.


{\bf Acknowledgments}\\
This work was supported under Contract No. MOST 107-2112-M-008
-023MY2
\mbox{}\\
E-mail address: mtkuo@ee.ncu.edu.tw\\

\setcounter{section}{0}

\renewcommand{\theequation}{\mbox{A.\arabic{equation}}} 
\setcounter{equation}{0} 

\mbox{}\\

{\bf Data Availability Statements}\\

The data that supports the findings of this study are available
within the article.

\end{document}